
\newtoks\fig
\message{You need PiCTeX and 256 kB TeX capacity for figures.}
\message{Do you want figures? (y/n)}
\read-1 to \fig
\if y\fig
  \message{Wait, wait, wait...}
  \input pictex.tex
  \long\def\figure#1\endfigure{\topinsert#1\endinsert}
\else
  \long\def\figure#1\endfigure{}
\fi
\magnification 1200
\hsize = 15 truecm \vsize = 22 truecm \hoffset = 1 truecm
\font\eightrm=cmr8 \font\eightit=cmti8 \font\eighti=cmmi8
\def\petit{\def\rm{\fam0\eightrm}%
           \textfont0=\eightrm \textfont1=\eighti
           \textfont\itfam=\eightit \def\it{\fam\itfam\eightit}
           \baselineskip=9pt \rm}
\def\sec#1{\par\goodbreak\medskip\noindent{\bf #1}\par\nobreak\medskip
\nobreak\noindent}
\def\e{\varepsilon}
\def\r{\rho}
\def\s{\sigma}
\def\z{\zeta}
\def\bni{\vphantom{\big)}}
\def\T{\Theta}\def\Tq{\T_q}\def\Tqq{\T_{q^2}}
\def\half{{1\over2}}
\def\[#1;#2]{(#1;#2)_\infty}
\def\<#1>{\[#1;q,x^4]}
\def\Tr{\,{\rm Tr}\ }
\def\snh{\,{\rm snh}\ }
\def\sn{\,{\rm sn}\ }
\def\G#1{\Gamma\left(#1\right)}
\def\clines#1{\vcenter{\openup1\jot\halign{\strut\hfil$\displaystyle##$\hfil
             \crcr#1\crcr}}}
\def\ar#1 #2 , #3 #4 {\arrow <0.3truecm> [0.1,0.3] from #1 #2 to #3 #4 }
\def\ln#1 #2 , #3 #4 {\arrow <0.01truecm> [0,0] from #1 #2 to #3 #4 }
\def\point#1 #2 {\put{\hbox{\kern -1pt .}} [Bl] at #1 #2 }
\def\bfpoint#1 #2 {%
\put{\hbox{\kern -2pt \raise -2pt%
\hbox{$\textstyle\bullet$}}} [Bl] at #1 #2 }
\rightline{LANDAU-94-TMP-5}
\rightline{hep-th/9408131}
\rightline{August 1994}
\vskip 3 truecm
\centerline{{\bf EQUATIONS FOR CORRELATION FUNCTIONS}}
\centerline{{\bf OF EIGHT-VERTEX MODEL:}}
\centerline{{\bf FERROMAGNETIC AND DISORDERED PHASES}%
\footnote{$^*$}{This work was supported in part by
Internatinal Science Foundation, International Association
for Promotion of Collaboration with Scientists of Former Soviet
Union, and Russian Foundation of Fundumental Investigations.}}
\vskip 1 truecm
\centerline{M. Yu. Lashkevich}
\vskip 1 truecm
\centerline{Landau Institute for Theoretical Physics, Kosygina 2, GSP-1,}
\centerline{117940 Moscow V-334, Russia}
\centerline{{\it E-mail: lashkevi@cpd.landau.free.net}}
\vskip 2 truecm
{\petit
The Kyoto group (Jimbo, Miwa, Nakayashiki {\it et al}\/.) showed that
the partition function and correlation functions of the eight-vertex
model in antiferromagnetic phases can be calculated using simple
analytical properties of the $R$-matrix. We extend these methods
to ferromagnetic and disordered phases. We use Baxter's symmetries
to obtain appropriate parametrizations of the $R$-matrix and to
substantiate the validity of the analytical approach for these phases.
These symmetries allow one to relate correlation functions in
different phases.}
\vfill\eject\insertpenalties=1000
\sec{1. Introduction}%
Exactly solvable lattice models have numerous applications to statistical
physics, quantum field theory and solid state physics. Effective methods
for calculating partition functions based on analytical properties were
proposed by Baxter.$^1$ Recently, the Kyoto group (Jimbo, Miwa,
Nakayashiki, \dots) considerably simplified these methods and
developed an approach to calculating correlation functions at finite
distances.$^{2-8}$ They applied this approach to the antiferromagnetic
phases of the six-vertex$^{2,3}$ and
eight-vertex$^{7,8}$ models and to the SOS and RSOS models.$^{4-6}$
An interesting challenge is to generalize this approach
to ferromagnetic and disordered phases of the eight-vertex (and
six-vertex) model. The problem is that the usual arguments for
applicability of the corner transfer matrix approach break down
in these phases. On the other hand, Baxter found some symmetries
between different phases of the eight-vertex model.$^1$
More precisely, the partition function is invariant
under some transformations of local wieghts, and these transformations
connect values of weights corresponding to different stable phases.
We show that these symmetries can be extended to correlation functions.
Then one can express correlation functions in ferromagnetic and
disordered phases in terms of correlators in the antiferromagnetic
phase. We formulate the rules for calculating correlation functions
directly in the ferromagnetic and disordered phases using appropriate
parametrizations of the local weights.

Now we state the notations. `Spins' of the eight-vertex
model $\e=\pm\equiv\pm1$ are situated at the links. The interaction
is attached to the vertices and described by the weight matrix
(or $R$-matrix) $R_{\e_1\e_2}^{\e_3\e_4}$ (Fig. 1a)
$$
R=\bordermatrix{&++&+-&-+&--\cr
++&a&&&d\cr+-&&b&c&\cr-+&&c&b&\cr--&d&&&a}.
\eqno(1.1)
$$
The partition function is given by (see Fig. 2)
$$
Z=\sum_{\{\mu_{kl},\nu_{kl}\}}\prod_{k,l\in{\bf Z}}
R_{\mu_{k-1,l}}^{\mu_{kl\vphantom{,}}}{}_{\nu_{kl}}^{\nu_{k,l-1}}.
$$
We need one more element (Fig. 1b). It is the spin flop
operator $\s^1=\pmatrix{&1\cr1&}$. It may be considered as a kind
of a two-link vertex.

\figure
%
%
\null\vskip 1 truecm
\hbox{
\beginpicture
\setcoordinatesystem units <1truecm,1truecm> point at 0 0
\ln 2.5 2 , 2.5 1
\ln 3 1.5 , 2 1.5
\put{$R_{\e_1\e_2}^{\e_3\e_4}=$} [Bl] at 0.2 1.4
\put{$\e_1$} [Bl] at 2.6 0.9
\put{$\e_2$} [Bl] at 1.9 1.7
\put{$\e_3$} [Bl] at 2.6 1.9
\put{$\e_4$} [Bl] at 3   1.6
\put{$(a)$} [Bl]  at 1.6 -0.2
\ln 6 1.5 , 7 1.5
\ln 6.4 1.4 , 6.6 1.6
\ln 6.6 1.4 , 6.4 1.6
\put{$(\s^1)_{\e_1}^{\e_2}=$} [Bl] at 4.2 1.4
\put{$\e_1$} [Bl]  at 5.9 1.7
\put{$\e_2$} [Bl]  at 6.8 1.7
\put{$(b)$} [Bl]   at 5.7 -0.2
\setcoordinatesystem units <1truecm,1truecm> point at -9 0
\ar 4 1 , 4 4
\ar 4 1 , 0.5 1
\ln 2 2 , 2 3
\ln 1.5 2.5 , 2.5 2.5
\ln 4 2.5 , 4.2 2.5
\ln 2 1 , 2 0.8
\put{$k$} [Bl]           at 4.3 2.4
\put{$l$} [Bl]           at 1.9 0.3
\put{$\mu_{kl}$} [Bl]    at 1.7 3.2
\put{$\mu_{k-1,l}$} [Bl] at 1.7 1.7
\put{$\nu_{kl}$} [Bl]    at 0.8 2.5
\put{$\nu_{k,l-1}$} [Bl] at 2.2 2.7
\endpicture
}

\nobreak\medskip\nobreak\noindent
\hskip 3 truecm Fig. 1\hskip 7 truecm Fig. 2
\endfigure

There are five regions of parameters$^1$ $a$, $b$, $c$, $d$
$$
\eqalign{
A_1:
&\quad c>a+b+d\qquad\hbox{(main region)},
\cr
A_2:
&\quad d>a+b+c,
\cr
D:
&\quad\half(a+b+c+d)>a,b,c,d,
\cr
F_1:
&\quad a>b+c+d,
\cr
F_2:
&\quad b>a+c+d.
}\eqno(1.2)
$$
The regions $A_1$ and $A_2$ are antiferromagnetic,
$F_1$ and $F_2$ ferromagnetic, and $D$ disordered.
The case $d=0$ corresponds to the six-vertex model.

The solution of the eight-vertex model in the antiferromagnetic
region is based on appropriate parametrization of the $R$-matrix
so that it satisfies some functional equations. We recall main
points of this approach in Sec. 2.
A simple local transformation makes it possible
to find necessary parametrization
in the ferromagnetic case (Sec. 3). Some nonlocal transformation
(`duality') connects the ferromagnetic region with the disordered one
(Sec. 4). A particular case of the six-vertex model
is considered in Sec. 5. Several concluding remarks are given in Sec. 6.
\sec{2. Antiferromagnetic Region}%
In this section we recall main points of the Kyoto approach. Consider the
antiferromagnetic region $A_1$. Let  $a_1^{(i)}$, $i=0,1$ be
ground state configurations
$$
\mu_{kl}=\nu_{kl}=(-)^{k+l+i}.
\eqno(2.1)
$$
Let $A_1^{(i)}$ be the set of configurations that differ from the
ground state $a_1^{(i)}$ by a finite number of spin flops. Only
configurations from $A_1^{(0)}\cup A_1^{(1)}$ contribute to the
partition function.

The $R$-matrix
can be parametrized as follows$^{1,7}$
$$
\eqalign{
a(\z)
&\equiv a(\z|q,x)=\r(\z)\cdot{\z\over x}
\Tqq(qx^2)\Tqq(q\z^2)\Tqq(x^2/\z^2),
\cr
b(\z)
&\equiv b(\z|q,x)=\r(\z)\cdot{1\over\z}
\Tqq(qx^2)\Tqq(\z^2)\Tqq(qx^2/\z^2),
\cr
c(\z)
&\equiv c(\z|q,x)=\r(\z)\cdot{1\over x}
\Tqq(x^2)\Tqq(q\z^2)\Tqq(qx^2/\z^2),
\cr
d(\z)
&\equiv d(\z|q,x)=\r(\z)\cdot{q^\half\over x^2}
\Tqq(x^2)\Tqq(\z^2)\Tqq(x^2/\z^2).
}\eqno(2.2)
$$
Here
$$
\clines{
\T_p(z)=\[z;p]\[p/z;p]\[p;p],
\cr
\[z;p_1,\cdots,p_N]=\prod_{n_1,\cdots,n_N=0}^\infty
\left(1-zp_1^{n_1}\cdots p_N^{n_N}\right).
}$$
The function $\T_p(z)$ is connected with the standart
theta-function $\T(u)$ as follows
$$
\T(iu)=\Tqq\left(qe^{-\pi u/I}\right),\quad q=e^{-\pi I'/I},
$$
with $I$, $I'$ being the standard halfperiods.

The region $A_1$ corresponds to the values of parameters
$$
0<q^\half<x<\z<1.
\eqno(2.3)
$$
The limit $q\rightarrow0$ corresponds to the six-vertex model.

The $R$-matrix satisfies the Yang--Baxter equation$^1$
$$
\eqalign{
\sum_{\alpha\beta\gamma}R(\z_1/\z_2)_{\e_1}^\alpha{}_{\e_2}^\beta
R(\z_1/\z_3)_\alpha^{\e_4}{}_{\e_3}^\gamma
&R(\z_2/\z_3)_\beta^{\e_5}{}_\gamma^{\e_6}
\cr
&=\sum_{\alpha'\beta'\gamma'}
R(\z_2/\z_3)_{\e_2}^{\beta'}{}_{\e_3}^{\gamma'}
R(\z_1/\z_3)_{\e_1}^{\alpha'}{}_{\gamma'}^{\e_6}
R(\z_1/\z_2)_{\alpha'}^{\e_4}{}_{\beta'}^{\e_5}.
}\eqno(2.4)
$$
Grafically this equation is presented in Fig. 3. Spectral parameters
$\z_i$ are attached to oriented lines.

\figure
%
%
\null\vskip 1 truecm
\hbox{
\beginpicture
\setcoordinatesystem units <1truecm,1truecm> point at 0 0
\ar 1   3 , 1   0.5
\ar 2.5 2 , 0.5 0.75
\ar 2.5 1 , 0.5 2.5
\put{$\z_1$} [Bl] at 0.9 0
\put{$\z_2$} [Bl] at 0.1 0.4
\put{$\z_3$} [Bl] at 0.1 2.7
\put{$\e_1$} [Bl] at 1.2 0.5
\put{$\e_2$} [Bl] at 0.3 1
\put{$\e_3$} [Bl] at 0.5 2
\put{$\e_4$} [Bl] at 1.2 2.6
\put{$\e_5$} [Bl] at 2.1 2.1
\put{$\e_6$} [Bl] at 2.1 0.7
\put{$=$} [Bl] at 3.3 1.4
\ar 6   3   , 6   0.5
\ar 6.5 2.5 , 4.5 1.25
\ar 6.5 1   , 4.5 2.5
\put{$\z_1$} [Bl] at 5.9 0
\put{$\z_2$} [Bl] at 4.1 0.9
\put{$\z_3$} [Bl] at 4.1 2.5
\put{$\e_1$} [Bl] at 5.5 0.7
\put{$\e_2$} [Bl] at 4.9 1.2
\put{$\e_3$} [Bl] at 4.9 2.4
\put{$\e_4$} [Bl] at 5.5 2.5
\put{$\e_5$} [Bl] at 6.5 2
\put{$\e_6$} [Bl] at 6.4 1.2
\endpicture
}

\nobreak\medskip\nobreak\noindent
\hskip 3 truecm Fig. 3
\endfigure

\figure
%
%
\null\vskip 1 truecm
\hbox{
\beginpicture
\setcoordinatesystem units <1truecm,1truecm> point at 0 0
\ar 1   2   , 1   1
\ar 1.5 1.5 , 0.5 1.5
\put{$\z$} [Bl] at 0.9 0.5
\put{$\z$} [Bl] at 0.2 1.4
\put{$=$} [Bl] at 1.8 1.4
\ar 3   2   , 2.5 1.5
\ar 3.5 1.5 , 3   1
\put{$\z$} [Bl] at 2.7 0.6
\put{$\z$} [Bl] at 2.2 1.1
\put{$(a)$} [Bl] at 1.8 -0.2
\ln 6.5 1.75 , 5.5 1.75
\ar 5.5 1.75 , 5.5 0.75
\ln 6   2.25 , 6   1.25
\ar 6   1.25 , 5   1.25
\put{$\z_1$} [Bl] at 5.4 0.35
\put{$\z_2$} [Bl] at 4.6 1.15
\put{$=$} [Bl] at 6.8 1.4
\ar 8.5 2.25 , 7.5 1.25
\ar 9   1.75 , 8   0.75
\put{$\z_1$} [Bl] at 7.6 0.35
\put{$\z_2$} [Bl] at 7.1 0.85
\put{$(a)$} [Bl] at 6.8 -0.2
\ar 11   2   , 11   1
\ar 11.5 1.5 , 10.5 1.5
\put{$\z_1$} [Bl] at 10.9 0.6
\put{$\z_2$} [Bl] at 10.1 1.2
\put{$\e_1$} [Bl] at 11.1 0.9
\put{$\e_2$} [Bl] at 10.4 1.7
\put{$\e_3$} [Bl] at 11.1 1.9
\put{$\e_4$} [Bl] at 11.4 1.6
\put{$=$} [Bl] at 11.8 1.4
\ar 13   2   , 13   1
\ar 12.5 1.5 , 13.5 1.5
\put{$\z_1$} [Bl]  at 12.8 0.6
\put{$x\z_2$} [Bl] at 13.8 1.4
\put{$-\e_1$} [Bl] at 13.1 0.9
\put{$\e_2$} [Bl]  at 12.4 1.7
\put{$-\e_3$} [Bl] at 12.8 2
\put{$\e_4$} [Bl]  at 13.4 1.6
\put{$(c)$} [Bl] at 11.8 -0.2
\endpicture
}

\nobreak\medskip\nobreak\noindent
\hskip 6.5 truecm Fig. 4
\endfigure

The function $\r(\z)$ in Eq. (2.2) is chosen so that the partition function
`per cite' is equal to unity. It is equivalent to the conditions$^7$
(Fig. 4)
$$
\eqalignno{
R(1)_{\e_1\e_2}^{\e_3\e_4}
=\delta_{\e_1}^{\e_4}\delta_{\e_2}^{\e_3}
&\qquad\hbox{(initial condition)},
&(2.5{\rm a})
\cr
\sum_{\alpha\beta}R(\z)_{\e_1}^\alpha{}_{\e_2}^\beta
R(\z^{-1})_\beta^{\e_3}{}_\alpha^{\e_4}
=\delta_{\e_1}^{\e_4}\delta_{\e_2}^{\e_3}
&\qquad\hbox{(unitarity)},
&(2.5{\rm b})
\cr
R(\z)_{\e_1\e_2}^{\e_3\e_4}
=R(x/\z)_{\e_4,-\e_1}^{\e_2,-\e_3}
&\qquad\hbox{(crossing symmetry)}.
&(2.5{\rm c})
}$$
The last condition (2.5c) is in a sense the crucial one.
It is not connected with normalization. Unnormalized but obviously
analytical $R$-matrix $\r^{-1}(\z)\,R(\z)$ satisfies Eq. (2.5c).
Therefore corresponding partition function `per cite' $\r^{-1}(\z)$
satisfies the equation
$$
\r(x/\z)=\r(\z).
\eqno(2.5{\rm c}')
$$
It should be mentionned that in the region $A_1$ the crossing
(2.5c) does not change boundary conditions (2.1).
Moreover, going from $\z$ in (2.3) to $x/\z$ we do not
intersect the surfaces of phase transitions. Indeed, if $0<x<\z<1$
then $0<x<x/\z<1$. But within one region the partition function per cite
$\r^{-1}(\z)$ is analytical and the condition (2.5c$'$)
is valid as an equation for an analytical function $\r(\z)$.
It is just Eq. (2.5c) for the parametrization (2.2)
that breaks down in the ferromagnetic and disordered
regions. Indeed, in these regions the right hand side and the
left hand side correspond to different phases with different
greatest eigenvalues of the transfer matrix.
The partition function per cite $\r^{-1}(\z)$
for unnormalized analytical weights
cannot be analytically continued over the phase transition.
Hence, the parametrization (2.2) is useless in ferromagnetic
and disordered phases.

The solution to Eqs. (2.5) is given by$^{1,7}$
$$
\eqalign{
\r(\z)
&=x\[q;q]^{-2}\[q^2;q^2]^{-1}\[x^2\z^{-2};q]^{-1}
\[qx^{-2}\z^2;q]^{-1}
\cr
&\times{\<x^4\z^2>\<x^2\z^{-2}>\<q\z^2>\<qx^2\z^{-2}>
\over\<x^4\z^{-2}>\<x^2\z^2>\<q\z^{-2}>\<qx^2\z^2>}.
}\eqno(2.6)
$$

We now turn to correlation functions. The Kyoto approach uses two
main types of objects: corner transfer matrices (CTMs)
and vertex operators (VOs). The north-west
corner transfer matrix,$^1$ $C_{NW}^{(i)}(\z)_{\e_1\e_2\cdots}
^{\e'_1\e'_2\cdots}$, is the partition function on the north-west
quadrant over the set $A_1^{(i)}$ with the configuration
$\e_1\e_2\cdots$ on the vertical boundary and $\e'_1\e'_2\cdots$
on horizontal one (Fig. 5a). The definitions of the south-west,
south-east and north-east CTMs,
$C_{SW}^{(i)}(\z)$, $C_{SE}^{(i)}(\z)$ and $C_{NE}^{(i)}(\z)$, (Fig. 5b)
are evident. On the infinite lattice
$$
\eqalignno{
C_{NW}^{(i)}(\z)
&=\z^{D^{(i)}},
&(2.7{\rm a})
\cr
C_{SW}^{(i)}(\z)
&=\s_\infty^1(x/\z)^{D^{(i)}},
&(2.7{\rm b})
\cr
C_{SE}^{(i)}(\z)
&=\s_\infty^1\z^{D^{(i)}}\s_\infty^1,
&(2.7{\rm c})
\cr
C_{NE}^{(i)}(\z)
&=(x/\z)^{D^{(i)}}\s_\infty^1,
&(2.7{\rm d})
}$$
$$
C_{NE}^{(i)}(\z)C_{SE}^{(i)}(\z)C_{SW}^{(i)}(\z)C_{NW}^{(i)}(\z)
=x^{2D^{(i)}},
\eqno(2.7{\rm e})
$$
where $D^{(i)}$ is an operator independent of $\z$, and
$$
\s_\infty^1=\s^1\otimes\s^1\otimes\cdots
$$
is a spin flop operator along one boundary of a CTM. Note that Eqs.
(2.7b--d) follow from Eq. (2.7a) and the crossing symmetry (2.5c).
It means that they only hold for the antiferromagnetic region.

\figure
%
%
\null\vskip 1 truecm
\hbox{
\beginpicture
\setcoordinatesystem units <1truecm,1truecm> point at 0 0
\ar 3   3.7 , 3   1
\ar 2.5 3.2 , 2.5 1
\ar 2   2.7 , 2   1
\ar 1.5 2.2 , 1.5 1
\ar 1   1.7 , 1   1
\ar 3.3 3.5 , 2.6 3.5
\ar 3.3 3   , 2.1 3
\ar 3.3 2.5 , 1.6 2.5
\ar 3.3 2   , 1.1 2
\ar 3.3 1.5 , 0.6 1.5
\put{$\z_1$} [Bl] at 0.9 0.5
\put{$\z_2$} [Bl] at 0.2 1.4
\put{$\e_1$} [Bl] at 3.4 1.5
\put{$\e_2$} [Bl] at 3.4 2
\point 3.5 2.4 \point 3.5 2.8 \point 3.5 3.2
\put{$\e'_1$} [Bl] at 3   0.7
\put{$\e'_2$} [Bl] at 2.5 0.7
\point 2.3 0.8 \point 1.9 0.8 \point 1.5 0.8
\put{$\z=\z_1/\z_2$} [Bl] at 0.2 3.3
\put{$(a)$} [Bl] at 1.8 -0.2
\ar 5.5 3.2 , 5.5 2.1
\ar 6   3.7 , 6   1.6
\ar 6.5 4.2 , 6.5 1.1
\ar 7   4.2 , 7   1.1
\ar 7.5 3.7 , 7.5 1.6
\ar 8   3.2 , 8 2.1
\ar 7.2 4   , 6.1 4
\ar 7.7 3.5 , 5.6 3.5
\ar 8.2 3   , 5.1 3
\ar 8.2 2.5 , 5.1 2.5
\ar 7.7 2   , 5.6 2
\ar 7.2 1.5 , 6.1 1.5
\setdashes
\ln 6.75 4.5  , 6.75 0.7
\ln 4.8  2.75 , 8.5  2.75
\setsolid
\put{$C_{NW}^{(i)}(\z)$} [Bl] at 4.2 3.8
\put{$C_{SW}^{(i)}(\z)$} [Bl] at 4.2 1.2
\put{$C_{SE}^{(i)}(\z)$} [Bl] at 7.5 1.2
\put{$C_{NE}^{(i)}(\z)$} [Bl] at 7.5 3.8
\put{$(b)$} [Bl] at 6.4 -0.2
\ar 13   4.2 , 13   1
\ar 13.5 2.5 , 12.5 2.5
\ar 13.5 2   , 12.5 2
\ar 13.5 1.5 , 12.5 1.5
\point 13.3 2.9 \point 13.3 3.3 \point 13.3 3.7
\point 12.7 2.9 \point 12.7 3.3 \point 12.7 3.7
\put{$\z_1$} [Bl] at 12.7 0.6
\put{$\z_2$} [Bl] at 12.1 1.1
\put{$\e_1$} [Bl] at 13.5 1.5
\put{$\e_2$} [Bl] at 13.5 2
\point 13.6 2.4 \point 13.6 2.7 \point 13.6 3
\put{$\e'_1$} [Bl] at 12.2 1.5
\put{$\e'_2$} [Bl] at 12.2 2
\point 12.3 2.4 \point 12.3 2.7 \point 12.3 3
\put{$\e$} [Bl] at 13.2 1
\put{$\z=\z_1/\z_2$} [Bl] at 12.2 4.3
\endpicture
}

\nobreak\medskip\nobreak\noindent
\hskip 4.5 truecm Fig. 5\hskip 7 truecm Fig. 6
\endfigure

The vertex operator
$$
\Phi_\e^{(1-i,i)}(\z)_{\e_1\e_2\cdots}^{\e'_1\e'_2\cdots}
=\sum_{\{\mu_k\}}\prod_{k=1}^\infty
R(\z)_{\mu_{k-1}\vphantom{'}}^{\mu_k}{}_{\e'_k}^{\e_k}\Big|_{\mu_0=\e},
\quad\e_k=-\e'_k=(-)^{k+i+1},\quad k\gg0.
\eqno(2.8)
$$
is a partition function along a half line (Fig. 6).
VOs satisfy the equations
$$
\sum_{\e'_1\e'_2}R(\z_1/\z_2)_{\e_1\e_2}^{\e'_1\e'_2}
\Phi_{\e'_1}^{(1-i,i)}(\z_1)\,\Phi_{\e'_2}^{(i,1-i)}(\z_2)
=\Phi_{\e_2}^{(1-i,i)}(\z_2)\,\Phi_{\e_1}^{(i,1-i)}(\z_1),
\eqno(2.9{\rm a})
$$
$$
\eqalignno{
\xi^{D^{(1-i)}}\Phi_\e^{(1-i,i)}(\z)
&=\Phi_\e^{(1-i,i)}(\z/\xi)\,\xi^{D^{(i)}},
&(2.9{\rm b})
\cr
\s_\infty^1\Phi_\e^{(1-i,i)}(\z)
&=\Phi_{-\e}^{(i,1-i)}(\z)\,\s_\infty^1.
&(2.9{\rm c})
}$$
Two first eqations follow from the Yang--Baxter equation (2.4),
Eq. (2.9c) is evident.

Every correlation function can be expressed in terms of vertex operator
correlation functions
$$
F_{\e_1\cdots\e_n}(\z_1,\cdots,\z_n)
=\Tr\left(x^{2D^{(i)}}\Phi_{\e_1}^{(i,1-i)}(\z_1)
\cdots\Phi_{\e_n}^{(1-i,i)}(\z_n)\right),\qquad
n=0,2,4,\cdots.
\eqno(2.10)
$$
Indeed, consider, for example, the probability $P_{\e_1\cdots\e_m}$
that $m$ spins along a row of parallel links take the values
$\e_1,\cdots,\e_m$ (Fig. 7a) in the phase $A_1^{(i)}$. Then
$$
\eqalign{
P_{\e_1\cdots\e_m}
&=\Tr\left(C_{NE}^{(i')}(\z)C_{SE}^{(i')}(\z)
\Phi_{\e_m}^{(i',1-i')}(\z)\cdots\Phi_{\e_1}^{(1-i,i)}(\z)\right.
\cr
&\left.\times C_{SW}^{(i)}(\z)C_{NW}^{(i)}(\z)
\Phi_{\e_1}^{(i,1-i)}(\z)\cdots\Phi_{\e_m}^{(1-i',i')}(\z)\right)
\cr
&=F_{-\e_m\cdots-\e_1\e_1\cdots\e_m}^{(i)}
(\underbrace{x\z,\cdots,x\z}_m,\underbrace{\z,\cdots,\z}_m).
}\eqno(2.11)
$$
The last line is obtained using Eqs. (2.9b) and (2.9c).
In general case it is necessary to divide the crystal (Fig. 7b)
into the CTMs 1, products of VOs 2, and a finite cluster
of $R$-matrices 3.

\figure
%
%
\null\vskip 1 truecm
\hbox{
\beginpicture
\setcoordinatesystem units <1truecm,1truecm> point at 0 0
\ln 5.2 3.5 , 4   3.5
\setdashes
\ln 4   3.5 , 3   3.5
\setsolid
\ar 3   3.5 , 1.1 3.5
\ln 5.7 3   , 4   3
\setdashes
\ln 4   3   , 3   3
\setsolid
\ar 3   3   , 0.6 3
\ln 5.7 2.5 , 4   2.5
\setdashes
\ln 4   2.5 , 3   2.5
\setsolid
\ar 3   2.5 , 0.6 2.5
\ln 5.2 2   , 4   2
\setdashes
\ln 4   2   , 3   2
\setsolid
\ar 3   2   , 1.1 2
\ar 1   3.2 , 1   2.1
\ar 1.5 3.7 , 1.5 1.6
\ar 2   3.7 , 2   1.6
\ar 2.5 3.7 , 2.5 1.6
\ar 4.5 3.7 , 4.5 1.6
\ar 5   3.7 , 5   1.6
\ar 5.5 3.2 , 5.5 2.1
\bfpoint 2   2.75
\bfpoint 2.5 2.75
\bfpoint 4.5 2.75
\setdashes
\ln 6   2.75 , 5   2.75
\ln 4.5 2.75 , 3   2.75
\ln 2   2.75 , 0.2 2.75
\ln 1.75 4   , 1.75 1.2
\ln 2.25 4   , 2.25 3
\ln 2.25 2.5 , 2.25 1.2
\ln 2.75 4   , 2.75 3
\ln 2.75 2.5 , 2.75 1.2
\ln 4.25 4   , 4.25 1.2
\ln 4.75 4   , 4.75 3
\ln 4.75 2.5 , 4.75 1.2
\setsolid
\put{$C_{NW}^{(i)}(\z)$} [Bl] at -0.1 4.2
\put{$C_{SW}^{(i)}(\z)$} [Bl] at -0.1 1.2
\put{$C_{SE}^{(i)}(\z)$} [Bl] at  5.5 1.2
\put{$C_{NW}^{(i)}(\z)$} [Bl] at  5.5 4.2
\put{$\Phi_{\e_1}^{(i)}(\z)$} [Bl] at 1.7 0.5
\put{$\Phi_{\e_1}^{(i)}(\z)$} [Bl] at 1.7 4.5
\put{$\Phi_{\e_m}^{(i)}(\z)$} [Bl] at 4.2 0.5
\put{$\Phi_{\e_m}^{(i)}(\z)$} [Bl] at 4.2 4.5
\put{$\e_1$} [Bl] at 2.1 2.7
\put{$\e_2$} [Bl] at 2.6 2.7
\put{$\e_m$} [Bl] at 4.6 2.7
\put{$(a)$} [Bl] at 2.7 -0.2
\ln  9.5 1.7 ,  9.5 3.7
\ln 10   1.3 , 10   4.2
\ln 10.5 1.3 , 10.5 4.2
\ln 11   1.3 , 11   4.2
\ln 11.5 1.3 , 11.5 4.2
\ln 12   1.3 , 12   4.2
\ln 12.5 1.7 , 12.5 3.7
\ln  9.8 1.5 , 12.2 1.5
\ln  9.3 2   , 12.7 2
\ln  9.3 2.5 , 12.7 2.5
\ln  9.3 3   , 12.7 3
\ln  9.3 3.5 , 12.7 3.5
\ln  9.8 4   , 12.2 4
\setdashes
\ln 10.25 1 , 10.25 4.5
\ln 11.75 1 , 11.75 4.5
\ln  9 2.25 , 13 2.25
\ln  9 3.25 , 13 3.25
\setsolid
\bfpoint 9.75 3.75
\ln 9.75 3.75 , 9 4.5
\ln 9 4.5 , 8.5 4.5
\put{1} [Bl] at 8.6 4.6
\bfpoint 9.75 2.75
\ln 9.75 2.75 , 8.5 2
\ln 8.5 2 , 8 2
\put{2} [Bl] at 8.1 2.1
\bfpoint 10.75 2.75
\ln 10.75 2.75 , 11.5 5
\ln 11.5 5 , 12 5
\put{3} [Bl] at 11.6 5.1
\put{1. CTMs} [Bl]        at 13.5 4
\put{2. VOs} [Bl]         at 13.5 3.5
\put{3. $R$-matrices} [Bl] at 13.5 3
\put{$(b)$} [Bl] at 11.7 -0.2
\endpicture
}

\nobreak\medskip\nobreak\noindent
\hskip 6.5 truecm Fig. 7
\endfigure

VO correlation functions can be found by solving the equations$^7$
$$
\clines{
F_{\e_1\cdots\e_n}^{(i)}(\xi\z_1,\cdots,\xi\z_n)
=F_{\e_1\cdots\e_n}^{(i)}(\z_1,\cdots,\z_n),
\cr
\sum_{\e'_j\e'_{j+1}}
R(\z_j/\z_{j+1})_{\e_j\e_{j+1}}^{\e'_j\e'_{j+1}}
F_{\cdots\e'_j\e'_{j+1}\cdots}^{(i)}(\cdots,\z_j,\z_{j+1},\cdots)
=F_{\cdots\e_{j+1}\e_j\cdots}^{(i)}(\cdots,\z_{j+1},\z_j,\cdots),
\cr
F_{\e_1\e_2\cdots\e_n}^{(i)}(x^2\z_1,\z_2,\cdots,\z_n)
=F_{\e_2\cdots\e_n\e_1}^{(1-i)}(\z_2,\cdots,\z_n,\z_1),
\cr
\sum_\e F_{-\e\e\e_1\cdots\e_n}^{(i)}(x\z,\z,\z_1,\cdots,\z_n)
=F_{\e_1\cdots\e_n}^{(i)}(\z_1,\cdots,\z_n),
\cr
F^{(i)}()=1.
}\eqno(2.12)
$$
For example,$^7$
$$
{\sum_\e\e F_{-\e\e}^{(0)}(\z,1)
\over\sum_\e F_{-\e\e}^{(0)}(\z,1)}
={\[x^3\z^{-1};x^2]\[x\z;x^2]\[-x\z^{-1};q]\[-qx^{-1}\z;q]
\over\[-x^3\z^{-1};x^2]\[-x\z;x^2]\[x\z^{-1};q]\[qx^{-1}\z;q]},
$$
and the Baxter--Kelland formula$^1$ for the `staggered' spontaneous
polarization holds
$$
\langle\e\rangle={\[x^2;x^2]^2\[-q;q]^2\over\[-x^2;x^2]^2\[q;q]^2}.
\eqno(2.13)
$$
\sec{3. Ferromagnetic Region}%
Consider the eight-vertex model in the main region $A_1$.
Consider the substitution
$$
\mu_{kl}=(-)^{k+l}\mu'_{kl},\quad\nu_{kl}=(-)^{k+l}\nu'_{kl}.
\eqno(3.1)
$$
Introduce new $R$-matrix, $R_F$, so that
$$
R_{\mu_{k-1,l}}^{\mu_{kl\vphantom{,}}}{}_{\nu_{kl}}^{\nu_{k,l-1}}
=(R'_F)_{\mu'_{k-1,l}}^{\mu'_{kl\vphantom{,}}}
{}_{\nu'_{kl}}^{\nu'_{k,l-1}}
=(R'_F)_{\mu_{k-1,l}}^{-\mu_{\vphantom{,}kl}}
{}_{-\nu_{kl}}^{\nu_{k,l-1}},
\eqno(3.2{\rm a})
$$
or, in components,
$$
a_F=c,\quad b_F=d,\quad c_F=a,\quad d_F=b.
\eqno(3.2{\rm b})
$$
Evidently
$$
a_F>b_F,c_F,d_F,
$$
and the matrix $R_F$ describes the model in the ferromagnetic $F_1$
region. The ground state antiferromagnetic configuration $a_1^{(i)}$
transforms to a ground state ferromagnetic configuration $f_1^{(i)}$
$$
\mu_{kl}=\nu_{kl}=(-)^i
\eqno(3.3)
$$
under the substitution (3.1). Similarly, the set $A_1^{(i)}$
transforms to the set $F_1^{(i)}$ of configurations that differ
from $f_1^{(i)}$ by finite number of spin flops.

The substitution (3.1) defines a natural parametrization in the
ferromagnetic region
$$
\eqalign{
a_F(\z)
&=\r(\z)\cdot{1\over x}\Tqq(x^2)\Tqq(q\z^2)\Tqq(qx^2/\z^2),
\cr
b_F(\z)
&=\r(\z)\cdot{q^\half\over x^2}\Tqq(x^2)\Tqq(\z^2)\Tqq(x^2/\z^2),
\cr
c_F(\z)
&=\r(\z)\cdot{\z\over x}\Tqq(qx^2)\Tqq(q\z^2)\Tqq(x^2/\z^2),
\cr
d_F(\z)
&=\r(\z)\cdot{1\over\z}\Tqq(qx^2)\Tqq(\z^2)\Tqq(qx^2/\z^2),
}\eqno(3.4)
$$
with $q$, $x$ and $\z$ in the region (2.3). The $R$-matrix $R_F(\z)$
satisfies the Yang--Baxter equation (2.4) and equations (2.5a,b).
The crossing symmetry is different (Fig. 8)
$$
R_F(\z)_{\e_1\e_2}^{\e_3\e_4}=R_F(x/\z)_{\e_4\e_1}^{\e_2\e_3}.
\eqno(3.5)
$$

\figure
%
%
\null\vskip 1 truecm
\hbox{
\beginpicture
\setcoordinatesystem units <1truecm,1truecm> point at 0 0
\ar 1   2   , 1   1
\ar 1.5 1.5 , 0.5 1.5
\put{$\z_1$} [Bl] at 0.9 0.4
\put{$\z_2$} [Bl] at 0.1 1.2
\put{$\e_1$} [Bl] at 1.1 0.9
\put{$\e_2$} [Bl] at 0.4 1.7
\put{$\e_3$} [Bl] at 1.1 1.9
\put{$\e_4$} [Bl] at 1.4 1.6
\put{$=$} [Bl] at 1.8 1.4
\ar 3   2   , 3   1
\ar 2.5 1.5 , 3.5 1.5
\put{$\z_1$} [Bl]  at 2.8 0.4
\put{$x\z_2$} [Bl] at 3.8 1.4
\put{$\e_1$} [Bl]  at 3.1 0.9
\put{$\e_2$} [Bl]  at 2.4 1.7
\put{$\e_3$} [Bl]  at 3.1 1.9
\put{$\e_4$} [Bl]  at 3.4 1.6
\endpicture
}

\nobreak\medskip\nobreak\noindent
\hskip 1.5 truecm Fig. 8
\endfigure

\noindent
Then the CTMs are given by
$$
\eqalign{
&C_{NW}^{(i)}(\z)=C_{SE}^{(i)}(\z)=\z^{D^{(i)}},
\cr
&C_{SW}^{(i)}=C_{NE}^{(i)}(\z)=(x/\z)^{D^{(i)}}
}\eqno(3.6)
$$
with some new $D^{(i)}$. The VOs, $\Phi_\e^{(i)}(\z)$ satisfy
the equations
$$
\clines{
\sum_{\e'_1\e'_2}R_F(\z_1/\z_2)_{\e_1\e_2}^{\e'_1\e'_2}
\Phi_{\e'_1}^{(i)}(\z_1)\,\Phi_{\e'_2}^{(i)}(\z_2)
=\Phi_{\e_2}^{(i)}(\z_2)\,\Phi_{\e_1}^{(i)}(\z_1),
\cr
\xi^{D^{(i)}}\Phi_\e^{(i)}(\z)=\Phi_\e^{(i)}(\z/\xi)\,\xi^{D^{(i)}}.
}\eqno(3.7)
$$
These equations do not relate objects with different values of $i$
and we omit the superscript $^{(i)}$ from now on if it does not lead
to a confusion. The VO correlation
functions
$$
F_{\e_1\cdots\e_n}(\z_1,\cdots,\z_n)
=\Tr\left(x^{2D}\Phi_{\e_1}(\z_1)\cdots\Phi_{\e_n}(\z_n)\right),
\qquad n=0,1,2,\cdots,
\eqno(3.8)
$$
satisfy the equations
$$
\clines{
F_{\e_1\cdots\e_n}(\xi\z_1,\cdots,\xi\z_n)
=F_{\e_1\cdots\e_n}(\z_1,\cdots,\z_n),
\cr
\sum_{\e'_j\e'_{j+1}}\!
R_F(\z_j/\z_{j+1})_{\e_j\e_{j+1}}^{\e'_j\e'_{j+1}}
F_{\cdots\e'_j\e'_{j+1}\cdots}(\cdots,\z_j,\z_{j+1},\cdots)
=F_{\cdots\e_{j+1}\e_j\cdots}(\cdots,\z_{j+1},\z_j,\cdots),
\cr
F_{\e_1\e_2\cdots\e_n}^{(i)}(x^2\z_1,\z_2,\cdots,\z_n)
=F_{\e_2\cdots\e_n\e_1}^{(1-i)}(\z_2,\cdots,\z_n,\z_1),
\cr
\sum_\e F_{\e\e\e_1\cdots\e_n}^{(i)}(x\z,\z,\z_1,\cdots,\z_n)
=F_{\e_1\cdots\e_n}^{(i)}(\z_1,\cdots,\z_n),
\cr
F()=1,\qquad F_\e^{(i)}(\z)=\delta_{\e,(-1)^i}.
}\eqno(3.9)
$$
The probabilities $P_{\e_1\cdots\e_m}$ are given by
$$
P_{\e_1\cdots\e_m}=F_{\e_m\cdots\e_1\e_1\cdots\e_m}
(\underbrace{x\z,\cdots,x\z}_m,\underbrace{\z,\cdots,\z}_m).
\eqno(3.10)
$$
The substitution (3.1) identifies Eqs. (3.9), (3.10) with
Eqs. (2.12), (2.11) for even $n$. It proves the correctness
of the above reasoning. In particular
$$
F_{\e_1\e_2\cdots\e_{n-1}\e_n}^{(i)\ {\rm fer.}}
(\z_1,\z_2,\cdots,\z_{n-1},\z_n)
=F_{-\e_1\e_2\cdots-\e_{n-1}\e_n}^{(i)\ {\rm antifer.}}
(\z_1,\z_2,\cdots,\z_{n-1},\z_n).
\eqno(3.12)
$$

To conclude this section we shall make several remarks.

Firstly, if one does not want to work with equations for VO correlation
functions, one may use the parametrization (2.2), but with new
value of the parameter $x$ and new function $\r(\z)$. The connection
between the parametrizations (2.2) and (3.4) is given in the Appendix.

The six-vertex limit $d_F\rightarrow0$ corresponds [in
parametrization (3.4)!] to
$$
q\rightarrow0,\qquad-q^{-\half}x={\rm const}=x_F.
\eqno(3.12)
$$
Here $x_F$ is the parameter of the six-vertex model corresponding
to $x$ in the parametrization (2.2).

The spontaneous polarization in the ferromagnetic phase is given
by the same formula (2.13) (but with the relation (3.4) between
$\z$, $q$, $x$ and weights!). In the six-vertex limit
(3.12) we obtain $\langle\e\rangle=1$. This result is consistent
with the wellknown fact that the ferromagnetic phase in
the six-vertex model is frozen.

Another interesting question is the sense of VO correlation
functions with odd $n$. Physically they correspond to a crystal
with dislocations. It is easy to check, for example, that the
spin at the link of a single dislocation for the eight-vertex
model in the ferromagnetic phase
is frozen $\langle\e\rangle=1$. There is an analogue
to these correlation functions in the antiferromagnetic region:
correlation functions with one insertion of $\s_\infty^1$.
\sec{4. Disordered Region}%
The duality transformation$^1$
$$
\eqalign{
&a_D=\half(a_F+b_F+c_F+d_F),
\cr
&b_D=\half(a_F+b_F-c_F-d_F),
\cr
&c_D=\half(a_F-b_F+c_F-d_F),
\cr
&d_D=\half(a_F-b_F-c_F+d_F)
}\eqno(4.1)
$$
connects the disordered region with the ferromagnetic region $F_1$.
Physically the duality is based on the representation of
the $R$-matrix
$$
\eqalign{
&(R_F)_{\e_1\e_2}^{\e_3\e_4}
=(R_1)_{\e_1\e_2}^{\e_3\e_4}+\cdots+(R_8)_{\e_1\e_2}^{\e_3\e_4},
\cr
&\eqalign{
(R_1)_{\e_1\e_2}^{\e_3\e_4}
&=\half a_D,
\cr
(R_3)_{\e_1\e_2}^{\e_3\e_4}
&=\half b_D\e_1\e_3,
\cr
(R_5)_{\e_1\e_2}^{\e_3\e_4}
&=\half c_D\e_1\e_4,
\cr
(R_7)_{\e_1\e_2}^{\e_3\e_4}
&=\half d_D\e_3\e_4,
}\qquad
\eqalign{
(R_2)_{\e_1\e_2}^{\e_3\e_4}
&=\half a_D\e_1\e_2\e_3\e_4,
\cr
(R_4)_{\e_1\e_2}^{\e_3\e_4}
&=\half b_D\e_2\e_4,
\cr
(R_6)_{\e_1\e_2}^{\e_3\e_4}
&=\half c_D\e_2\e_3,
\cr
(R_8)_{\e_1\e_2}^{\e_3\e_4}
&=\half d_D\e_1\e_2.
}}\eqno(4.2)
$$
The partition function can be written as
$$
Z=\sum_{\{t_{kl}=1,\cdots,8\}}\sum_{\{\mu_{kl},\nu_{kl}\}}
\prod_{kl}\left(R_{t_{kl}}\right)
_{\mu_{k-1,l}}^{\mu_{kl}}{\phantom{)}}_{\nu_{kl}}^{\nu_{k,l-1}}.
$$
Let us sum up over all $\mu_{kl}$, $\nu_{kl}$. A nonzero contribution
into the sum over, for example, $\mu_{kl}$ is given by the terms
$$
\sum_{\mu_{kl}}\left(R_{t_{kl\vphantom{,}}}\right)
_{\mu_{k-1,l}}^{\mu_{kl\vphantom{,}}}{\bni}_{\nu_{kl}}^{\nu_{k,l-1}}
\left(R_{t_{k+1,l}}\right)
_{\mu_{kl}}^{\mu_{k+1,l}}{\bni}_{\nu_{k+1,l}}^{\nu_{k+1,l-1}}
\eqno(4.3)
$$
with such $t_{kl}$ and $t_{k+1,l}$ that the power of $\mu_{kl}$
is even. Now one can associate a dual
vertex $\matrix{\e_3&\e_4\cr\e_1&\e_2}$ to any $R_t$
(Fig. 9) and a dual spin to any pair $(t_{kl},t_{k+1,l})$ or
$(t_{kl},t_{k,l-1})$ that gives a nonzero contribution:
$$
\eqalign{
&R_1\rightarrow\matrix{+&+\cr+&+},
\cr
&R_5\rightarrow\matrix{+&-\cr-&+},
}\quad
\eqalign{
&R_2\rightarrow\matrix{-&-\cr-&-},
\cr
&R_6\rightarrow\matrix{-&+\cr+&-},
}\quad
\eqalign{
&R_3\rightarrow\matrix{-&+\cr-&+},
\cr
&R_7\rightarrow\matrix{-&-\cr+&+},
}\quad
\eqalign{
&R_4\rightarrow\matrix{+&-\cr+&-},
\cr
&R_8\rightarrow\matrix{+&+\cr-&-}
}\eqno(4.4{\rm a})
$$
\figure
%
%
\null\vskip 1 truecm
\hbox{
\beginpicture
\setcoordinatesystem units <1truecm,1truecm> point at 0 0
\put{$R_1$} [Bl] at 0.8 2
\ln 0.5 1 , 1.5 1
\ln 1 0.5 , 1 1.5
\put {$+$} [Bl] at 1.1 0.4
\put {$+$} [Bl] at 0.4 1.1
\put {$+$} [Bl] at 1.1 1.3
\put {$+$} [Bl] at 1.3 1.1
\setcoordinatesystem units <1truecm,1truecm> point at -1.5 0
\put{$R_2$} [Bl] at 0.8 2
\ln 0.5 1 , 1.5 1
\ln 1 0.5 , 1 1.5
\put {$-$} [Bl] at 1.1 0.4
\put {$-$} [Bl] at 0.4 1.1
\put {$-$} [Bl] at 1.1 1.3
\put {$-$} [Bl] at 1.3 1.1
\setcoordinatesystem units <1truecm,1truecm> point at -3 0
\put{$R_3$} [Bl] at 0.8 2
\ln 0.5 1 , 1.5 1
\ln 1 0.5 , 1 1.5
\put {$-$} [Bl] at 1.1 0.4
\put {$+$} [Bl] at 0.4 1.1
\put {$-$} [Bl] at 1.1 1.3
\put {$+$} [Bl] at 1.3 1.1
\setcoordinatesystem units <1truecm,1truecm> point at -4.5 0
\put{$R_4$} [Bl] at 0.8 2
\ln 0.5 1 , 1.5 1
\ln 1 0.5 , 1 1.5
\put {$+$} [Bl] at 1.1 0.4
\put {$-$} [Bl] at 0.4 1.1
\put {$+$} [Bl] at 1.1 1.3
\put {$-$} [Bl] at 1.3 1.1
\setcoordinatesystem units <1truecm,1truecm> point at -6 0
\put{$R_5$} [Bl] at 0.8 2
\ln 0.5 1 , 1.5 1
\ln 1 0.5 , 1 1.5
\put {$-$} [Bl] at 1.1 0.4
\put {$+$} [Bl] at 0.4 1.1
\put {$+$} [Bl] at 1.1 1.3
\put {$-$} [Bl] at 1.3 1.1
\setcoordinatesystem units <1truecm,1truecm> point at -7.5 0
\put{$R_6$} [Bl] at 0.8 2
\ln 0.5 1 , 1.5 1
\ln 1 0.5 , 1 1.5
\put {$+$} [Bl] at 1.1 0.4
\put {$-$} [Bl] at 0.4 1.1
\put {$-$} [Bl] at 1.1 1.3
\put {$+$} [Bl] at 1.3 1.1
\setcoordinatesystem units <1truecm,1truecm> point at -9 0
\put{$R_7$} [Bl] at 0.8 2
\ln 0.5 1 , 1.5 1
\ln 1 0.5 , 1 1.5
\put {$+$} [Bl] at 1.1 0.4
\put {$+$} [Bl] at 0.4 1.1
\put {$-$} [Bl] at 1.1 1.3
\put {$-$} [Bl] at 1.3 1.1
\setcoordinatesystem units <1truecm,1truecm> point at -10.5 0
\put{$R_8$} [Bl] at 0.8 2
\ln 0.5 1 , 1.5 1
\ln 1 0.5 , 1 1.5
\put {$-$} [Bl] at 1.1 0.4
\put {$-$} [Bl] at 0.4 1.1
\put {$+$} [Bl] at 1.1 1.3
\put {$+$} [Bl] at 1.3 1.1
\endpicture
}

\nobreak\medskip\nobreak\noindent
\hskip 5.5 truecm Fig. 9
\endfigure
\noindent
for vertices,
$$
\eqalign{
(t,t')\rightarrow+\qquad
&\hbox{if}\qquad t=1,4,5,8,\qquad t'=1,4,6,7,
\cr
(t,t')\rightarrow-\qquad
&\hbox{if}\qquad t=2,3,6,7,\qquad t'=2,3,5,8
}\eqno(4.4{\rm b})
$$
for vertical links, and
$$
\eqalign{
(t,t')\rightarrow+\qquad
&\hbox{if}\qquad t=1,3,6,8,\qquad t'=1,3,5,7,
\cr
(t,t')\rightarrow-\qquad
&\hbox{if}\qquad t=2,4,5,7,\qquad t'=2,4,6,8
}\eqno(4.4{\rm c})
$$
for horizontal ones. Then the summation over dual spins is
equivalent to the summation over $\{t_{kl}\}$, and this dual
model corresponds to the local weights $a_D$, $b_D$, $c_D$, $d_D$.
This is the Baxter's construction of duality.$^1$ Note that the
dual lattice here coincides with the initial one and is not
shifted by a half period as it is for the Ising model.

Now we shall extend this duality to correlation functions.
Let us insert into the lattice several spin flop operators
$\s^1$. If, for example, one of them is inserted into the
vertical link $kl$, the expression (4.3) turns into
$$
\eqalign{
\sum_{\mu_{kl}}\left(R_{t_{kl\vphantom{,}}}\right)
_{\mu_{k-1,l}}^{\mu_{kl\vphantom{,}}}{\bni}_{\nu_{kl}}^{\nu_{k,l-1}}
&\left(R_{t_{k+1,l}}\right)
_{-\mu_{kl}}^{\mu_{k+1,l}}{\bni}_{\nu_{k+1,l}}^{\nu_{k+1,l-1}}
\cr
&=\pm\sum_{\mu_{kl}}\left(R_{t_{kl\vphantom{,}}}\right)
_{\mu_{k-1,l}}^{\mu_{kl\vphantom{,}}}{\bni}_{\nu_{kl}}^{\nu_{k,l-1}}
\left(R_{t_{k+1,l}}\right)
_{\mu_{kl}}^{\mu_{k+1,l}}{\bni}_{\nu_{k+1,l}}^{\nu_{k+1,l-1}},
}
$$
where the sign is `$+$' if there is no $\mu_{kl}$ in both $R_{t_{kl}}$
and $R_{t_{k+1,l}}$ and `$-$' if $\mu_{kl}$ is there. From Eq. (4.4b) we
see that the sign coincides with the value of dual spin
at this link. It means that a spin flop operator in the initial lattice
corresponds to a spin variable in the dual lattice. Conversely,
a spin variable in the initial lattice
corresponds to a spin flop operator in the dual lattice.
It means that there is a relation between correlation functions
in the disordered and ferromagnetic regions
$$
\eqalign{
&\langle\mu_{k_1l_1}\cdots\mu_{k_ml_m}
\nu_{k'_1l'_1}\cdots\nu_{k'_nl'_n}
\s^1_{K_1L_1}\cdots\s^1_{K_ML_M}
\tau^1_{K'_1L'_1}\cdots\tau^1_{K'_NL'_N}\rangle_D
\cr
&=\half\sum_{i=0}^1\langle\s^1_{k_1l_1}\cdots\s^1_{k_ml_m}
\tau^1_{k'_1l'_1}\cdots\tau^1_{k'_nl'_n}
\mu_{K_1L_1}\cdots\mu_{K_ML_M}
\nu_{K'_1L'_1}\cdots\nu_{K'_NL'_N}\rangle_{F_1}^{(i)},
}\eqno(4.5)
$$
where $\s^1_{kl}$ and $\tau^1_{kl}$ are spin flop operators
$\s^1$ inserted into the vertical and
horizontal links $kl$ respectively. The summation over the boundary
conditions $i=0,1$ is necessary, because the construction
of duality needs summation over all configurations.

We now turn to parametrization. The transformation (4.1)
leads to the parametrization
$$
\clines{
\eqalign{
a_D(\z)
&={\cal N}\r(\z)\Tq(-\z)\Tq(-x/\z)\Tq(q^\half\z)\Tq(q^\half x/\z)
/\Tq(-x)\Tq(q^\half x),
\cr
b_D(\z)
&={\cal N}\r(\z)\Tq(\z)\Tq(x/\z)\Tq(-q^\half\z)\Tq(-q^\half x/\z)
/\Tq(-x)\Tq(q^\half x),
\cr
c_D(\z)
&={\cal N}\r(\z)\Tq(-\z)\Tq(x/\z)\Tq(q^\half\z)\Tq(-q^\half x/\z)
/\Tq(x)\Tq(-q^\half x),
\cr
d_D(\z)
&={\cal N}\r(\z)\Tq(\z)\Tq(-x/\z)\Tq(-q^\half\z)\Tq(q^\half x/\z)
/\Tq(x)\Tq(-q^\half x),
}\cr
{\cal N}={1\over x}{\[q;q]\[x^2;q]\[qx^{-2};q]\over\[-1;q]\[q^\half;q]^2}.
}\eqno(4.6)
$$
The $R$-matrix $R_D(\z)$ satisfies Eqs. (2.4), (2.5a,b) and the
same crossing symmetry equation as the ferromagnetic $R$-matrix
$$
R_D(\z)_{\e_1\e_2}^{\e_3\e_4}=R_D(x/\z)_{\e_4\e_1}^{\e_2\e_3}.
\eqno(4.7)
$$
So the CTMs are given by
$$
\eqalign{
&C_{NW}(\z)=C_{SE}(\z)=\z^D,
\cr
&C_{SW}(\z)=C_{NE}(\z)=(x/\z)^D,
}\eqno(4.8)
$$
and the VO, $\Phi_\e(\z)$, satisfies the equations
$$
\clines{
\sum_{\e'_1\e'_2}R_D(\z_1/\z_2)_{\e_1\e_2}^{\e'_1\e'_2}
\Phi_{\e'_1}(\z_1)\Phi_{\e'_2}(\z_2)
=\Phi_{\e_2}(\z_2)\Phi_{\e_1}(\z_1),
\cr
\xi^D\Phi_\e(\z)=\Phi_\e(\z/\xi)\xi^D.
}\eqno(4.9)
$$
Therefore the VO correlation functions
$$
F_{\e_1\cdots\e_n}(\z_1,\cdots,\z_n)
=\Tr\left(x^{2D}\Phi_{\e_1}(\z_1)\cdots\Phi_{\e_n}(\z_n)\right),
\qquad n=0,1,2,\cdots,
\eqno(4.10)
$$
satisfy the equations
$$
\clines{
F_{\e_1\cdots\e_n}(\xi\z_1,\cdots,\xi\z_n)
=F_{\e_1\cdots\e_n}(\z_1,\cdots,\z_n)
\cr
\sum_{\e'_j\e'_{j+1}}\!R_D(\z_j/\z_{j+1})_{\e_j\e_{j+1}}^{\e'_j\e'_{j+1}}
F_{\cdots\e'_j\e'_{j+1}\cdots}(\cdots,\z_j,\z_{j+1},\cdots)
=F_{\cdots\e_{j+1}\e_j\cdots}(\cdots,\z_{j+1},\z_j,\cdots),
\cr
F_{\e_1\e_2\cdots\e_n}(x^2\z_1,\z_2,\cdots,\z_n)
=F_{\e_2\cdots\e_n\e_1}(\z_2,\cdots,\z_n,\z_1),
\cr
\sum_\e F_{\e\e\e_1\cdots\e_n}(x\z,\z,\z_1,\cdots,\z_n)
=F_{\e_1\cdots\e_n}(\z_1,\cdots,\z_n),
\cr
F()=1,\qquad F_\e(\z)=\half
}\eqno(4.11)
$$
with Eq. (3.10) for probabilities.

Let us relate this construction to the ferromagnetic case. Let
$$
\Phi_\e(\z)=\Phi'_+(\z)+\e\Phi'_-(\z).
\eqno(4.12)
$$
It is easy to check using Eq. (4.1) or (4.2) that
$$
\sum_{\e'_1\e'_2}
R_F(\z_1/\z_2)_{\e_1\e_2}^{\e'_1\e'_2}
\Phi'_{\e'_1}(\z_1)\Phi'_{\e'_2}(\z_2)
=\Phi'_{\e_2}(\z_2)\Phi'_{\e_1}(\z_1),
$$
Therefore $\Phi'_\e(\z)$ is the ferromagnetic VO. More precisely,
$$
\Phi'_\e(\z)\sim(\Phi_\e^{(0)}(\z)+\Phi_\e^{(1)}(\z))_{\rm ferromag.}.
\eqno(4.13)
$$
This imposes an additional condition on the VO correlation functions
$$
F_{\e_1\cdots\e_n}(\z_1,\cdots,\z_n)
=F_{-\e_1\cdots-\e_n}(\z_1,\cdots,\z_n).
\eqno(4.14)
$$
On the other hand it is straightforward to obtain Eq. (4.5)
from (4.10)---(4.14).

We ought to make a remark. We have implicitly supposed that
$a,b,c,d>0$. But to cover the whole region
$\half(a_D+b_D+c_D+d_D)>a_D,b_D,c_D,d_D>0$ we need
to consider negative values of $a,b,c,d$. In this case the
applicability of the above construction can be
justified by the fact that there are no phase transitions
in the disordered region. Hence, the partition function
and correlation functions may be analytically continued
from the subregion $a,b,c,d>0$.$^1$
\sec{5. Six-Vertex Model in Critical Region}%
One of the most interesting problems is to obtain a descripition
of the six-vertex model in the critical region which corresponds
to the disordered region of the eight-vertex model with $d=0$.
We could set $d_D(\z)=0$, but it is not convenient, because
the crossing transformation (4.7) does not preserve this condition.
Instead, we apply the transformation described in Sec. 3 to the disordered
region
$$
a'_D(\z)=c_D(\z),\quad b'_D(\z)=d_D(\z),\quad
c'_D(\z)=a_D(\z),\quad d'_D(\z)=b_D(\z).
\eqno(5.1)
$$
This transformation maps the disordered region onto itself.
The crossing simmetry of the form (2.5c)
$$
a'_D(x/\z)=b'_D(\z),\qquad c'_D(x/\z)=c'_D(\z),
\qquad d'_D(x/\z)=d'_D(\z)
$$
preserves the condition $d'_D(\z)=0$. On the other hand,
the equations for VO correlation functions and probabilities
are given by (2.12), (2.11) for these weights, but without
superscripts $^{(i)}$, $^{(1-i)}$, with the substitution
$R(\z)\rightarrow R'_D(\z)$, and with additional condition (4.14).

It is convenient to use here an additive spectral parameter instead of
multiplicative one. Change variables
$$
\z=e^{-\pi u'/2I},\qquad x=e^{-\pi\lambda'/2I}.
\eqno(5.2)
$$
It is well known$^1$ that
$$
a:b:c:d=\snh(\lambda'-u'):\snh u':\snh\lambda'
:k\snh\lambda'\snh u'\snh(\lambda'-u'),
$$
where $\snh z=-i\sn iz$, and $k$ is the module of the elliptic functions.
Consider the limit $q^\half\rightarrow-1$ ($k\rightarrow-1$). Then
$$
a:b:c:d=\tan(\lambda'-u'):\tan u':\tan\lambda'
:-\tan\lambda'\tan u'\tan(\lambda'-u').
$$
In this limit $a+b-c-d=0$ or $d'_D=0$. Introducing new variables
$$
\clines{
u=2u',\qquad\lambda=2\lambda',
\cr
0<u<\lambda<\pi,
}\eqno(5.3)
$$
we easily obtain
$$
\eqalign{
a'_D(u)
&=\r'(u)\sin(\lambda-u),
\cr
b'_D(u)
&=\r'(u)\sin u,
\cr
c'_D(u)
&=\r'(u)\sin\lambda,
\cr
d'_D(u)
&=0,
}\eqno(5.4)
$$
where the function $\r'(u)$ providing initial condition,
unitarity, and crossing symmetry can be adopted from Ref. 9:
$$
\clines{
\r'(u)={1\over\pi}\G{\lambda\over\pi}\G{1-{\lambda\over\pi}}
\G{1-{\lambda-u\over\pi}}
\prod_{p=1}^\infty{r_p(u)r_p(\lambda-u)\over r_p(0)r_p(\lambda)},
\cr
r_p(u)={\G{2p\lambda-u\over\pi}\G{1+{2p\lambda-u\over\pi}}
\over\G{(2p+1)\lambda-u\over\pi}\G{1+{(2p-1)\lambda-u\over\pi}}}.
}\eqno(5.5)
$$
Now the equations for correlation functions in the critical region
of the six-vertex model are evident.
\sec{6. Conclusion}%
We extended the vertex operator formalism from the antiferromagnetic
region to the ferromagnetic and disordered ones. The crucial role
belongs to the pa\-ra\-me\-tri\-za\-tion
of the weights and to the crossing
symmetry equation which are different in different regions.
Baxter's symmetries prove the correctness of the result.

In this paper we did not consider vertex operators of type II.
These operators are very important because they diagonalize
explicitly the transfer matrix and give the spectrum of
exitations. Evidently, they can be introduced in every region
through the connection with the elliptic algebra
${\cal A}_{q,-x}(\widehat{sl}_2)$.$^8$
\sec{Acknoledgment}%
The author is grateful to T.~Miwa for a stimulating discussion.
\sec{\hfil Appendix\hfil}%
In practical calculations the relations between different
parametrizations may be useful. Here we cite them:
$$
\eqalign{
R_F(\z|q,x)
&=R(\z|q,x_F)|_{\r=\r_F},
\cr
R_D(\z|q,x)
&=R(\z_D|q_D,x_D)|_{\r=\r_D}.
}\eqno(A.1)
$$
Using the usual techniques of elliptic functions we obtain
$$
\clines{
x_F=-q^{-\half}x,
\cr
\r_F(\z|q,x_F)=-{q\z\over x_F^2}\r(\z|q,-q^\half x),
\cr
{1\over\log q_D}={1\over\log q}+{i\over 2\pi},
\quad q'_D=iq',
\quad\z_D=\z^\alpha,\quad x_D=(q^\half x)^\alpha,
\quad\alpha={\log q_D\over\log q},
\cr
{\r_D(\z_D|q_D,x_D)\over\r(\z|q,x)}
={1\over{\cal N}}{q_D^\half\over x_D^2}
{\T_{q_D^2}(x_D^2)\T_{q_D^2}(\z_D^2)\T_{q_D^2}(x_D^2/\z_D^2)
\Tq(x)\Tq(-q^\half x)
\over\Tq(\z)\Tq(-x/\z)\Tq(-q^\half\z)\Tq(q^\half x/\z)}.
}\eqno(A.2)
$$
Here the relation between the conjugate parameters
$q'$ and $q'_D$ is cited.

Note that the ferromagnetic spontaneous polarization can be written
in the form
$$
\langle\e\rangle={\[qx_F^2;qx_F^2]^2\[-q;q]^2
\over\[-qx_F^2;qx_F^2]^2\[q;q]^2}.
$$
\sec{References}%

\item{1.}R.~J.~Baxter, {\it Exactly Solved Models in Statistical
Mechanics}\/, Academic Press, London, 1982

\item{2.}B.~Davies, O.~Foda, M.~Jimbo, T.~Miwa and A.~Nakayashiki,
{\it Commun. Math. Phys.} {\bf 151}, 89 (1993)

\item{3.}M.~Jimbo, K.~Miki, T.~Miwa and A.~Nakayashiki,
{\it Phys. Lett.} {\bf A168}, 256 (1992)

\item{4.}M.~Jimbo, T.~Miwa and Y.~Ohta,
{\it Int. J. Mod. Phys.} {\bf A8}, 1457 (1993)

\item{5.}M.~Idzumi, T.~Tokihiro, K.~Iohara, M.~Jimbo, T.~Miwa
and J.~Nakashima, {\it Int. J. Mod. Phys.} {\bf A8}, 1479 (1993)

\item{6.}O.~Foda, M.~Jimbo, T.~Miwa, K.~Miki and A.~Nakayashiki,
{\it J. Math. Phys.} {\bf 35}, 13 (1994)

\item{7.}M.~Jimbo, T.~Miwa and A.~Nakayashiki,
{\it J. Phys.} {\bf A26}, 2199 (1993)

\item{8.}O.~Foda, K.~Iohara, M.~Jimbo, R.~Kedem, T.~Miwa and H.~Yan,
Research Institute for Mathematical Sciences preprint RIMS-979
(May 1994)

\item{9.}A.~B.~Zamolodchikov and Al.~B.~Zamolodchikov,
{\it Ann. Phys.} {\bf 120}, 253 (1979)

\end